\documentclass[twocolumn,prl,aps,amsmath,amssymb]{revtex4}

\usepackage{graphicx}
\usepackage{dcolumn}
\usepackage{bm}
\begin{document}
\title{\begin{flushright}
{\rm \small CERN--PH--TH/2007--068\\
SU--4252--8489\\}
\end{flushright}
\par \vskip .05in
Alternative Large $N_c$ Schemes and Chiral Dynamics }
 \author{Francesco {\sc Sannino}}\email{francesco.sannino@nbi.dk}
 \affiliation{ CERN Theory Division, CH-1211 Geneva 23, Switzerland.}
  \affiliation{ University of Southern Denmark, Campusvej 55, DK-5230,
 Odense M., Denmark.}

\author{Joseph {\sc Schechter}}\email{schechte@physics.syr.edu}
 \affiliation{Department of Physics, Syracuse University,
Syracuse, NY 13244-1130, USA.}
\date{March 2007}

\begin{abstract}
We compare the dependences on the number of colors of the
 leading $\pi\pi$ scattering
amplitudes using the single index quark
field and two index quark fields.
 These are seen to have
different relationships to the scattering amplitudes
 suggested by chiral dynamics
which can explain the long puzzling pion pion
s wave scattering up to about 1 GeV.
This may be interesting for getting a better understanding
of the large $N_c$ approach as well as for application to 
recently proposed technicolor models.

\end{abstract}

\maketitle

\section{Background}

     Gaining control of QCD in its strongly interacting
 (low energy) regime
 constitutes a real challenge. One very attractive approach
 is based on studying the theory in the large number
of colors ($N_c$) limit \cite{tH,Wi}. At the same time one may
obtain more information by requiring the theory to model the 
(almost) spontaneous breakdown of chiral
 symmetry \cite{njl,gl}. A standard test case is pion pion
scattering in the energy range up to about 1 GeV. Some time ago,
an attempt was made \cite{SS,HSS1} to implement this
combined scenario.
Since the leading large $N_c$ amplitude contains only tree
diagrams involving mesons of the standard quark-antiquark
type, it is expected that the required amplitude should be gotten by
calculating just the chiral tree diagrams for rho
 meson exchange together with
the four point pion contact diagram. There are no unknown
parameters in this calculation. The crucial question
is whether the scattering amplitude calculated in this way
will satisfy unitarity. When one 
compares the result
with experimental data up to about 1 GeV
on the real part of the (most sensitive
 to unitarity violation) J=I=0 partial wave, one finds 
(see Fig.1 of \cite{HSS1}) that the result violates the partial
wave unitarity bound by just a ``little bit". 
 On the other hand, the pion contact term by itself violates unitarity 
much more 
drastically so one might argue that the
 large $N_c$ approach, which suggests that the tree diagrams
of all quark anti-quark resonances in the relevant energy
range be included, is helping a 
lot. To make matters more quantitative one might ask the question:
by how much should $N_c$ be increased in order for the amplitude
in question to remain within the unitarity bounds
 for energies below 1 GeV?

\begin{figure}[htbp]
\includegraphics[width=
8.5cm,clip=true]{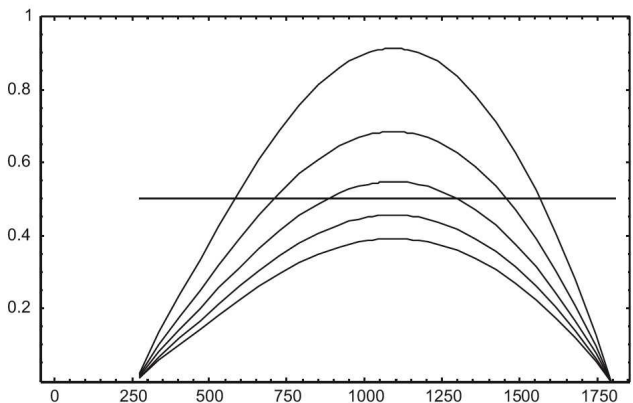}
\caption[]
{ Real part of the $I=J=0$ partial wave amplitude due
to the current algebra $+\rho$ terms, plotted for the following
increasing values of $N_c$ (from up to down),
$3,4,5,6,7$. The curve with largest
violation of the unitarity bound corresponds to $N_c=3$ while the
ones within the unitarity bound are for $N_c=6,7$.} \label{carho}
\end{figure}

    This question was answered in a very simple way in
 \cite{Harada:2003em}, as we now briefly review.
In terms of the conventional
amplitude, $A(s,t,u)$ the $I=0$ amplitude is
$3A(s,t,u)+A(t,s,u)+A(u,t,s)$. One gets the $J=0$ channel by
projecting out the correct partial wave.
The current algebra (pion contact diagram) contribution to the
conventional amplitude is
\begin{eqnarray}
A_{ca}(s,t,u)=2\frac{s-m^2_{\pi}}{F^2_{\pi}}\ ,
\label{eq:ca}
\end{eqnarray}
where the pion decay constant, $F_{\pi}$ depends on $N_c$ as
$F_{\pi}(N_c)=131\sqrt{N_c}/\sqrt{3}$ so
 that $F_{\pi}(3)=131$~MeV.
Furthermore $m_{\pi}=137$~MeV is independent of $N_c$.
 The desired amplitude
 is obtained by adding to the current algebra
term the following vector meson $\rho(770)$ contribution:
\begin{eqnarray}A_{\rho}(s,t,u)&=&
\frac{g_{\rho\pi\pi}^2}{2m^2_{\rho}}(4m^2_{\pi}-3s)\nonumber
\\&-&\frac{g_{\rho\pi\pi}^2}{2}\left[\frac{u-s}{(m^2_{\rho}-t)
-im_{\rho}\Gamma_{\rho}\theta(t-4m^2_{\pi})}
\right.\nonumber \\&+& \left.
\frac{t-s}{(m^2_{\rho}-u)-im_{\rho}\Gamma_{\rho}\theta(u-4m^2_{\pi})}\right]
\ ,\label{arho}
\end{eqnarray}
where $g_{\rho\pi\pi}(N_c)= 8.56
\sqrt{3}/\sqrt{N_c}$ is the $\rho \pi \pi$ coupling constant.
Also,
$m_{\rho}=771$~MeV is independent of $N_c$ and
 \begin{eqnarray}
\Gamma_{\rho}(N_c)=\frac{g^2_{\rho
\pi\pi}\left(N_c\right)}{12\pi\,m_{\rho}^2}
\left(\frac{m^2_{\rho}}{4} - m_{\pi}^2 \right)^{\frac{3}{2}} \ .
\end{eqnarray}            
It should be noted that the first term in Eq.~(\ref{arho}),
which is
an additional non-resonant contact interaction other than the current
algebra contribution, is required when we
include the $\rho$ vector meson contribution in a chiral invariant
manner.
In Fig.~\ref{carho} we show the real part of the $I=J=0$
amplitude (denoted $R^0_0$) due to current algebra plus the $\rho$
contribution for increasing values of $N_c$. Since in this channel
the vector meson is never on shell we suppress the contribution of
the width in the vector meson propagator in Eq.~(\ref{arho}).
One observes that the unitarity bound
 (i.e., $\vert R_0^0 \vert \le 1/2$)
 is satisfied for $N_c\geq 6$ till well beyond
the 1~GeV region. However unitarity is still a problem for $3,4$
and $5$ colors. {}At energy scales larger than the one associated
with the vector meson clearly other resonances are needed \cite{SS}
but we shall not be concerned with that energy range here. It is also
interesting to note that these considerations are essentially
unchanged when the pion mass (i.e. explicit chiral symmetry
 breaking in the Lagrangian) is set to zero.

    Note that essentially we are just using the scaling,
\begin{eqnarray}
A(s,t,u)=\frac{3}{N_c}\widetilde{A}(s,t,u) \ .
\label{Nc scaling}
\end{eqnarray}
where $\widetilde{A}(s,t,u)$ is defined replacing $F_{\pi}$ and
$g_{\rho\pi\pi}$ with the $N_c$ independent quantities
$\widetilde{F}_{\pi}= F_{\pi}\,\sqrt{3}/\sqrt{N_c}$ and
$\widetilde{g}_{\rho\pi\pi}=g_{\rho\pi\pi}\sqrt{N_c}/\sqrt{3}$.
Other authors \cite{oa} have found the same minimum
value, $N_c$=6 for the practical consistency of the large $N_c$
approximation, by using different methods.

In order to explain low energy $\pi\pi$ scattering for the physical
value
$N_c=3$ one must go beyond the large $N_c$ approximation.
It is attractive to keep the assumption
of tree diagram dominance involving near by resonances, however.
One easily sees that adding a scalar singlet resonance (sigma) at the 
location where the unitarity bound on 
$R_0^0(s)$
 is first violated
should restore unitarity. This is because the real part
 of a Breit Wigner
resonance is zero at the pole location and
negative just above it, so will
 bring $R_0^0(s)$ below the bound, as required.
 In \cite{Harada:2003em},
the resonance mass was found to be around 550 MeV on this basis.
 Such a low
value would make it different from a p-wave quark-antiquark state,
which is expected to be in the 1000-1400 MeV range.
We assume then that 
it is a four quark state (glueball states
 are expected to be in the 1.5
GeV range from lattice investigations).
 Four quark
states of diquark-anti diquark type \cite{Jaffe} and meson-meson type
\cite{molecule} have been
 discussed in the literature for many years.
Accepting this picture, however,  
poses a problem for the accuracy of the large $N_c$
inspired description of the scattering since
four quark states are  
predicted not to exist in the large $N_c$ limit of QCD.
We shall take the point of view that a four quark 
type state is present since it allows a natural fit to
the low energy data.
 Of course, it is still necessary to fine
tune the parameters and shape of the sigma resonance to fit
 the experimental $\pi \pi$ scattering data in detail.
 In practice, since the 
parameters
of the pion contact and rho exchange contributions are fixed, the 
sigma is the most important one for fitting and fits may 
even be achieved \cite{HSS2} if the vector meson
 piece is neglected. However
the well established, presumably four
 quark type, $f_0$(980) resonance
 must be included to achieve a fit in the region just
 around 1 GeV. 
 
 There is by now a fairly large recent literature
 \cite{kyotoconf}-\cite{b} 
on the effect of light ``exotic" scalars in low energy meson meson
 scattering. There seems to be a consenesus, arrived at using rather
different approaches (keeping however, unitarity), that the sigma exists.

\section{Two index quark fields}
      
 Now, consider redefining the $N_c=3$ quark field with
 color index A (and flavor
index not written) as
\begin{equation}
        q_A= \frac{1}{2} \epsilon_{ABC}q^{BC},{} q^{BC}=-q^{CB},
\label{redefine}
\end{equation}
so that, for example, $q_1=q^{23}$ and similarly for the adjoint field,
${\bar q}^1={\bar q}_{23}$ etc. This is just a
 trivial change of variables and will not 
change anything for QCD. However, if a continuation
 of the theory is made
to $N_c>3$ the resulting theory will be
 different since the two index
 antisymmetric quark representation has $N_c(N_c-1)/2$ rather
 than $N_c$ color components.
As was pointed out by Corrigan and Ramond
 \cite{Corrigan:1979xf}, who were
mainly interested in the problem of the
 baryons at large $N_c$, this shows
that the extrapolation of QCD to higher $N_c$ is not unique.
 Further investigation
of the properties of the alternative extrapolation model
introduced in \cite{Corrigan:1979xf} was 
carried out 
 by Kiritsis and Papavassiliou \cite{Kiritsis:1989ge}.
 Here, we shall discuss the 
consequences for the  low energy  $\pi \pi$ scattering
 discussed above, of this
alternative large $N_c$ extrapolation, assuming for our purpose,
 that all the quarks extrapolate as
 antisymmetric two index objects.   

It may be worthwhile to remark that gauge theories with
two index quarks have recently gotten a great deal of attention.
 Armoni, Shifman and Veneziano \cite{Armoni:2003gp,Armoni:2003fb,Armoni:2004ub,Armoni:2005wt,Armoni:2007rf} have
 proposed an interesting
 relation between certain sectors of the two index antisymmetric
 (and symmetric) theories at
 large number of colors and
sectors of super Yang-Mills (SYM).
Using a supersymmetry inspired
effective Lagrangian approach $1/N_c$ corrections were
investigated in \cite{hep-th/0309252}. Information on the super Yang-Mills spectrum has been obtained in \cite{Feo:2004mr}. On the validity of the large $N_c$ equivalence between different theories and interesting new possible phases we refer the reader to \cite{Unsal:2007fb,Unsal:2006pj,Kovtun:2005kh}. The finite temperature phase transition and its relation with chiral symmetry has been investigated in \cite{hep-th/0507251} while the effects of a nonzero
 baryon chemical potential were studied in \cite{Frandsen:2005mb}.

When adding flavors the phase
diagram as a function of the number of flavors and
 colors has been provided in \cite{hep-ph/0405209}. The  complete phase diagram for fermions in arbitrary representations has been unveiled in \cite{Dietrich:2006cm}. The study of theories with fermions in a higher dimensional
 representation of the gauge group and
 the knowledge of the associated conformal window led to
the construction of minimal models of technicolor
\cite{hep-ph/0405209,hep-ph/0406200, Dietrich:2005jn} which are not
ruled out by current precision measurements and lead to interesting dark matter candidates \cite{Kainulainen:2006wq,Gudnason:2006yj,Kouvaris:2007iq} as well as to a very high degree of unification of the standard model gauge couplings \cite{Gudnason:2006mk}.

Besides these two limits a third one for
 massless one-flavor QCD, which
 is in between the 't Hooft and Corrigan Ramond
ones, has been very recently
 proposed in \cite{Ryttov:2005na}.
 Here one first splits the
 QCD Dirac fermion into the two elementary Weyl fermions
 and afterwards assigns one of them
 to transform according to a rank-two antisymmetric
 tensor while the other
 remains in the fundamental representation of the
 gauge group. For three
 colors one reproduces
one-flavor QCD and for a generic number of colors the
theory is chiral. {}The generic $N_c$ is
 a particular case of the 
generalized
Georgi-Glashow model \cite{Georgi:1985hf}.

\begin{figure}[htbp]
\includegraphics[width=
7.5cm,clip=true]{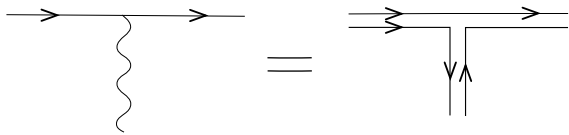}
\vspace*{ -7cm}
\caption[]
{Two index fermion - gluon vertex.} \label{FigA}
\end{figure}

   To illustrate the large $N_c$ counting for the $\pi\pi$ scattering
amplitude when quarks are designated
to transform according to the two index antisymmetric representation
of color SU(3) one may employ \cite{tH} the mnemonic where
each tensor index of this group is represented by a directed line.
Then the quark-quark gluon interaction is pictured as in Fig. \ref{FigA}. 
The two index quark is pictured as two lines with arrows pointing in the 
same direction, as opposed to the gluon which has two lines with arrows 
pointing in opposite directions. The coupling constant representing this 
vertex is taken to be $g_t/\sqrt{N_c}$, where $g_t$ (the 't Hooft 
coupling constant)  is to be held constant.

\begin{figure}[htbp]
\includegraphics[width=
3.0cm,clip=true]{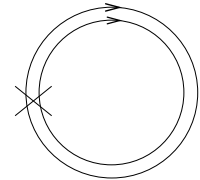}
\caption[]
{Diagram for $F_\pi$ for the two index quark.} \label{FigB}
\end{figure}

    A ``one point function", like the pion decay constant, $F_\pi$
would have as it's simplest diagram, Fig.~\ref{FigB}.
          
The X represents a pion insertion and is associated with a
normalization factor for the color part of the pion's wavefunction,
\begin{eqnarray}
            \frac{\sqrt{2}}{\sqrt{N_c(N_c-1)}},
\label{wf}
\end{eqnarray}        
which scales for large $N_c$ as $1/N_c$. The two circles each
carry a quark index so their joint factor scales as $N_c^2$ for
 large $N_c$; more
precisely, taking the antisymmetry into account, the factor is
\begin{equation}          
 \frac{N_c(N_c-1)}{2}.
\label{loop}
\end{equation}
The product of Eqs. (\ref{wf}) and (\ref{loop})
 yields the $N_c$ scaling for
$F_{\pi}$:
\begin{equation}
F_{\pi}^2(N_c)= \frac{N_c(N_c-1)}{6}F_{\pi}^2(3).
\label{fpiscaling}
\end{equation}
For large $N_c$, $F_\pi$ scales proportionately to $N_c$ rather
than to $\sqrt{N_c}$ as in the case of the 't Hooft extrapolation.

\begin{figure}[htbp]
\includegraphics[width=
4.0cm,clip=true]{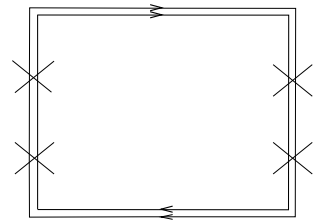}
\vspace*{ -1cm}
\caption[]
{Diagram for the scattering amplitude, A with the 2 index quark. } \label{FigC}
\end{figure}

    Using this scaling together with Eq.(\ref{eq:ca}) suggests that
 the $\pi \pi$ scatttering 
amplitude, $A$ scales as,
\begin{equation}
A(N_c)=\frac{6}{N_c(N_c-1)}A(3),
\label{Ascaling}
\end{equation}
which, for large $N_c$ scales as $1/N_c^2$ rather than
 as $1/N_c$  in
the 't Hooft extrapolation. This scaling law for
 large $N_c$ may be verified
from the mnemonic in Fig.~\ref{FigC}, where there is
 an $N_c^2$ factor from the two loops
multiplied by four factors of $1/N_c$ from the X's.

It is interesting to find the minimum value of $N_c$ for 
which the tree amplitude due to the pion and rho meson terms
(given in Eqs.(\ref{eq:ca}) and (\ref{arho}) above) is unitary in 
this two antisymmetric index quark extrapolation scheme.
 Fig.~\ref{carho} shows that the the peak value of 
the partial wave amplitude, $R_0^0$ due
to these two terms is numerically about
0.9. This is to be identified with $A_{ca}(3)+A_{\rho}(3)$ in
Eq.(\ref{Ascaling}). Thus the condition that the
 extrapolated amplitude be unitary is,
\begin{equation}
0.9 \frac{6}{N_c(N_c-1)} < 1/2.
\label{uncondition}
\end{equation}  
Clearly, the extrapolated amplitude is unitary already for
$N_c=4$, which indicates better convergence in $N_c$ than for
the 't Hooft case which became unitary at $N_c=6$.

\begin{figure}[htbp]
\includegraphics[width=
8.5cm,clip=true]{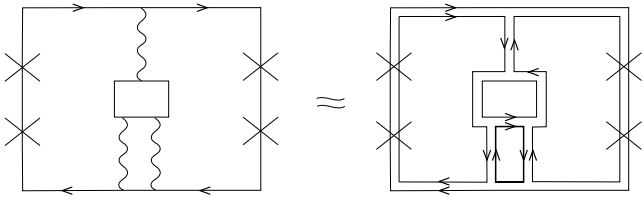}
\vspace*{ -3cm}
\caption[]
{Diagram for the scattering amplitude, A including an internal 2 index quark loop. } \label{FigD}
\end{figure}

    There is still another different feature; consider the typical
$\pi \pi$ scattering diagram with an extra internal (two index)
quark loop, as shown in Fig.~\ref{FigD}.

 In this diagram there are 
four X's (factor from Eq.(\ref{wf})), five index loops (factor 
from Eq.(\ref{loop})) and six gauge coupling constants. These
combine to give a large $N_c$ scaling behavior proportional
to $1/N_c^2$ for the $\pi \pi$ scattering amplitude. We see that
diagrams with an extra internal 2 index quark loop are not suppressed
compared to the leading diagrams. This is analogous,
as pointed out in \cite{Kiritsis:1989ge}, to the 
behavior of diagrams with an extra gluon loop in the 't Hooft
extrapolation scheme. Now, Fig.~\ref{FigD} is a diagram which can describe
a sigma particle exchange. Thus in the 2 index quark scheme,
``exotic" four quark resonances can appear at the leading order
in addition to the usual two quark resonances. Given the discussion 
we reviewed above, the possibility of a sigma appearing at leading order
means that one can construct a unitary $\pi \pi$ amplitude  
already at $N_c$ = 3 in the 2 antisymmetric index scheme. From the
 point of view
of low energy $\pi \pi$ scattering, it seems to be unavoidable to say
that the 2 index scheme is  more realistic than the 't Hooft
scheme 
given the existence of a four quark type sigma.

    Of course, the usual 't Hooft extrapolation has a number of
 other things to recommend it. These include the fact that nearly all
meson resonances seem to be of the quark- antiquark type, the OZI rule
predicted holds to a good approximation and baryons emerge
 in an elegant way as solitons in the model. 

 A fair statement would seem to be that each extrapolation emphasizes
different aspects of the true $N_c$ = 3 QCD.
 In particular, the usual scheme
is not really a replacement for the true theory. That appears to be the
 meaning of the fact that the continuation to $N_c>3$ is not unique.  

\section{Quarks in two index symmetric color representation}

    Clearly the assignment of quarks to the two index
 symmetric representation of
color SU(3) looks very similar.
 We may denote the quark fields
as,
 \begin{equation}
q_{AB}^{sym}=q_{BA}^{sym},
\label{symquarks}
\end{equation}
in contrast to Eq.(\ref{redefine}). There will be $N_c(N_c+1)/2$ 
different color states for the two index
 symmetric quarks. This means that
there is no value of $N_c$ for which
 the symmetric theory can be made to
 correspond to true QCD. For $N_c$ = 3 there
 are 6 color states of the quarks
and 8 color states of the gluon. If we
 choose $N_c$ = 2, there are 3 color states
of the quarks but unfortunately only three
 color states of the gluon.
 On the other hand, for large $N_c$ it would
 seem reasonable to make
approximations like,
 
\begin{equation}
A^{sym}(N_c) \approx A^{asym}(N_c),
\label{symasym}
\end{equation}
for the $\pi \pi$ scattering amplitude.

    As far as the large $N_c$ counting goes,
 the mnemonics in Figs. \ref{FigA}-\ref{FigD}
are still 
applicable to the case of quarks
 in the two index symmetric color
representation. For not so large $N_c$, the scaling
 factor for the pion insertion 
would be
\begin{eqnarray}
            \frac{\sqrt{2}}{\sqrt{N_c(N_c+1)}},
\label{symwf}
\end{eqnarray}        
and the pion decay constant would scale as
\begin{equation}
F_{\pi}^{sym}(N_c)\propto \sqrt{\frac{N_c(N_c+1)}{2}}.
\label{symfpiscaling}
\end{equation}

     With the identification $A^{QCD}=A^{asym}(3)$, the use of
Eq.(\ref{symasym})
enables us to estimate the large $N_c$
scattering amplitude as,
\begin{equation}
A^{sym}(N_c) \approx \frac{6}{N_c^2}A^{QCD}.
\label{bigN}
\end{equation}

In 
applications to recently proposed minimal walking technicolor theories
this formula 
is useful for making estimates involving weak gauge bosons via
the Goldstone boson equivalence theorem \cite{gbet}.

\begin{figure}[htbp]
\includegraphics[width=
4.5cm,clip=true]{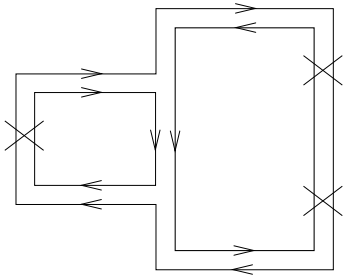}
\caption[]
{Diagram for meson decay into two glueballs.} \label{FigE}
\end{figure}

     Finally we remark on the large $N_c$ scaling rules for meson 
and glueball masses and decays in either the two index
antisymmetric or two index symmetric schemes. Both meson and
 glueball masses scale
as $(N_c)^0$. Furthermore, all six reactions of the type
\begin{equation}
a\rightarrow b + c,
 \label{abc}
\end{equation}
where a,b and c can stand for either a meson or a glueball,
scale as $1/N_c$. This is illustrated in Fig.\ref{FigE} for the case
of a meson decaying into two glueballs; note that the
glueball insertion scales as $1/N_c$ and that two interaction
 vertices are involved.

\vspace*{ -.5 cm}

\section{Summary}
We have investigated the dependences on the number of colors of the
 leading $\pi\pi$ scattering
amplitudes using the single and the two index quark fields.
 
We have seen that in the 2 index quark extension of QCD,
{\it exotic} four quark resonances can appear at the leading order
in addition to the usual two quark resonances. {}From the
 point of view of low energy $\pi \pi$ scattering
 the 2 index scheme is  more realistic than the 't Hooft one
 given the existence of a four quark type sigma. This allows
one to explain the long puzzling pion pion
s wave scattering up to about 1 GeV. 

 Of course, the usual 't Hooft extrapolation has a number of
 other important predictions to recommend it. A fair statement
 is that each large $N_c$ extrapolation of QCD captures 
different aspects of the physical $N_c$ = 3 case.

We have also related the QCD scattering amplitude
 at large $N_c$ with the one featuring two index symmetric
 quarks (Similar connections exist for adjoint fermions).
  The results are 
interesting
 for getting a better understanding
of the large $N_c$ approach as well as for application to 
recently proposed technicolor models.
 \acknowledgments
\section*{Acknowledgments}
\vspace*{ -.5cm} It is a pleasure to thank A. Abdel Rehim, D. Black, D.D. Dietrich,
A. H. Fariborz, M.T. Frandsen, M. Harada, S. Moussa, S. Nasri and K. 
Tuominen for
helpful discussions.  The work of F.S. is supported by the Marie Curie
Excellence Grant under contract MEXT-CT-2004-013510 as well as the Danish Research
Agency.
 The work of J.S. is supported in part by the U.
S. DOE under Contract no. DE-FG-02-85ER 40231.

\end{document}